\documentclass[preprint,authoryear,12pt]{elsarticle}

\usepackage{amssymb}
\usepackage{amsthm}

\usepackage{amsmath,amsthm,color}

\journal{Insurance: Mathematics and Economics}

\begin{document}

\begin{frontmatter}



\title{Optimal mean-variance investment strategy under value-at-risk constraints}


\author[jye]{Jun Ye\corref{jyecor}}

\cortext[jyecor]{Corresponding author. Tel.:+86-10-62788974.}

\address[jye]{Department of Mathematical Sciences, Tsinghua
University Beijing, 100084, China. }

\ead{jye@math.tsinghua.edu.cn}

\author[jye]{Tiantian Li \fnref{ltttel}}

\fntext[ltttel]{Tel.:+86-10-66931968}

\ead{litt08@mails.tsinghua.edu.cn}

\begin{abstract}
This paper is devoted to study the effects arising from imposing a
value-at-risk (VaR) constraint in mean-variance portfolio selection
problem for an investor who receives a stochastic cash flow which
he/she must then invest in a continuous-time financial
market. For simplicity, we assume that there is only one investment
opportunity available for the investor, a risky stock. Using
techniques of stochastic linear-quadratic (LQ) control, the optimal
mean-variance investment strategy with and without VaR constraint
are derived explicitly in closed forms, based on solution of
corresponding Hamilton-Jacobi-Bellman (HJB) equation. Furthermore,
some numerical examples are proposed to show how the addition of the
VaR constraint
affects the optimal strategy.\\

\end{abstract}


\end{frontmatter}


\section{Introduction}
The mean-variance model of \citet{Markowitz1952portselect,
Markowitz1959portselect} is a cornerstone
of modern portfolio theory. The most important contribution of this
model is that it enables an investor to optimally select
mean-variance efficient portfolios for seeking the highest return
after specifying his acceptable risk level. Since \citeauthor*{Markowitz1952portselect}'s
pioneering work, the mean-variance model was extended from
single-period case to multi-period discrete-time case \citep[see][etc.]{
Hakansson1971multi, Pliska1997intro, Samuelson1969lifetime}
and continuous time case during the last decades \citep[see][etc.]{Cox1989optimal,
Duffie1991mvhedging, Karatzas1987optimal, Schweizer1996approx}. However,
when studying these two kinds of dynamic portfolio selection models,
most research works have been dominated by those of maximizing
expected utility functions of the terminal wealth. Nevertheless,
when using this approach, the tradeoff information between the risk
and the expected return is implicit, which makes the investment
decision less intuitive. In \citeyear{xyzhou00lqframework}, \citeauthor*{xyzhou00lqframework} introduce the
stochastic linear-quadratic (LQ) control as a general framework to
study the mean-variance optimization problem. Within this framework
they have established a natural connection of the portfolio
selection problems and standard stochastic control models and
attained some elegant results for a continuous-time mean-variance
model with determined coefficients.

When using stochastic LQ control approach to deal with continuous
time mean-variance problem, the terminal wealth is a random variable
with a distribution that is often extremely skewed and shows
considerable probability in regions of small values of the terminal
wealth. This means that the optimal terminal wealth may exhibit
large shortfall risks. In order to prevent investors from extremely
dangerous positions in the market, it is thus more reasonable to
consider asymmetric risk measures, e.g. value-at-risk (VaR), to
limit the exposure to market risks.

In market risk management, it is widely accepted that VaR is a
useful summary measure of market risks which regulatory authorities
sometimes enforced investors to use. VaR is actually the maximum
expected loss over a given horizon period at a given level of
confidence. For comprehensive introduction to risk management using
VaR, we refer the reader to \citet{Jorion1997VaR}.

Recognizing that risk management is typically not an investor's
primary objective, the investors would like to limit their risks
while maximizing expected utility. This leads to stochastic control
problems under restrictions on such risk measures. There has been
considerable interest in the study of portfolio selection models
subject to a VaR constraint. \citet{Kluppelberg1998optimal},
\citet{Alexander2004compare, Alexander2007mvport}
investigate the optimal portfolio choices
subject to a VaR or conditional VaR (CVaR) constraint in a static
(one-period) setting. The similar problems in a dynamic setting has
started to draw more attentions recently, \citet{Basak2001VaRrisk}
focus on the optimal portfolio policies of a
utility-maximizing agent by imposing the VaR constraint at one point
in time. \citet{Cuoco2008optimaldyn} developed a realistic
dynamically consistent model of the optimal behavior of a trader
subject to risk constraints. They assume that the risk of the
trading portfolio is re-evaluated dynamically by using the
conditioning information, and hence the trader must satisfy the risk
limit continuously. \citet{Yiu2004optimal} explicitly derived the
standard VaR constraint on total wealth and obtained optimal trading
strategy(without consideration of re-insurance).
\cite{Pirvu2005max} started with the model of \citet{Cuoco2008optimaldyn}
and found the optimal growth portfolio subject
to these risk measures. \cite{Pirvu2007port} extended those
results by extensively studying the optimal investment and
consumption strategies for both logarithmic utility and
non-logarithmic CRRA utilities.

Motivated by \citet{xyzhou00lqframework} and \citet{Yiu2004optimal}, this
paper addresses the problem of an investor who receives an
uncontrollable stochastic cash flow which he must then invest in a
complete continuous-time financial market in order to maximize the
weighted average of the expectation and the variance of his terminal
wealth at a horizon time. The main focus in this paper is on the
mean-variance optimization problem of the investor subject to a risk
limit specified in terms of VaR on his future net worth. To our
knowledge, this problem has not yet received a complete treatment in
the existing literature. In this paper, we derive the optimal
mean-variance investment strategy under a standard VaR constraint by
solving the corresponding Hamilton-Jacobi-Bellman (HJB) equation and
explore how the addition of a risk constraint affects the optimal
solution.

The rest of the paper is organized as follows. Section 2 describes
the model, including the definition of the VaR on the future net
worth process. Section 3 contains the main characterization result
of the VaR constraint and formulates the portfolio optimization
problem that can be eventually discussed as a stochastic LQ problem.
Section 4 gives the explicit solution of the optimal mean-variance
strategy without VaR constraint by solving the corresponding HJB
equation. Section 5 discusses the optimal mean-variance strategy
with VaR constraint. Finally, Section 6 provides some numerical
examples to show how the addition of the VaR constraint affects the
optimal strategy.

\section{The Model}

 \subsection{Continuous-time investment in stochastic cash flow}
 All stochastic processes introduced below are supposed to be
adapted in a filtered probability space $(\Omega,\mathcal
{F},\mathcal {F}_t,P)$, where $\mathcal {F}_t, t\geq 0$ is a
filtration satisfying the usual conditions. Moreover, it is assumed
throughout this paper that all inequalities as well as equalities
hold $P$-almost surely.

Following the framework of \citet{brown95exputility}, for
simplicity, and without any loss of generality, we assume that there
is only one risky stock available for investment, whose price at
time $t$ will be denoted by $P_t$ which satisfies the following
stochastic differential equation
\begin{equation}\label{equ:stock_sde}
  dP_t=P_t(\mu dt+\sigma dW_t^{(1)}),
\end{equation}
where $\mu>0$ is the appreciation rate and $\sigma>0$ is the
volatility or the dispersion of the stock. $W_t^{(1)}$ is a standard
Brownian motion.

Since we are concerned with investment behavior in the presence of a
stochastic cash flow, or an external risk process denoted by $Y_t$,
which is another Browian motion with drift $\alpha$ and diffusion
parameter $\beta>0$, that is
\begin{equation}\label{equ:cashflow_sde}
  dY_t=\alpha dt+\beta dW_t^{(2)},
\end{equation}
where $W_t^{(1)}$ and $W_t^{(2)}$ are possibly correlated with
correlation coefficient $\rho$. In case there would only be one
source of randomness left in the model, we also assume that
$\rho^{2}<1$.

We will define an investment strategy $f$ as an admissible adapted
control process $f_t$, satisfying that $\int_0^T f_t^2 dt<\infty$,
a.s., for all $T<\infty$. Note that $f_t$ represents the amount
invested in risky stock at time $t$, and we will not put more
constraints on $f_t$. In particular, we will allow $f_t<0$, which
means short-selling would be allowed, the circumstance that
$f_t>X_t^f$ is also permitted so that the investor can borrow money
to buy stock.

Assume that the trading takes place continuously and transaction
cost is not considered. Therefore, following the investment strategy
$f$, the wealth process of the investor at time $t$ , which will be
denoted by $X_t^f$, can be given by the following stochastic
differential equation with initial condition $X_0=x$
\begin{equation}\label{equ:wealth_sde}
  dX_t=(f_t\mu+\alpha)dt+f_t\sigma dW_t^{(1)}+\beta dW_t^{(2)},X_0=x.
\end{equation}\\

 \subsection{Value-at-risk}
Now we want to introduce the definition of value-at-risk. Here we start
by rewriting (\ref{equ:wealth_sde}) into integration form
\begin{equation}\label{equ:wealth_int}
  X_t^f=x+\int_0^t(f_s\mu+\alpha)ds+\int_0^t(f_s\sigma)dW_s^{(1)}+\int_0^t\beta dW_s^{(2)},
\end{equation}
where $x>0$ denotes the initial value of the portfolio.
Notice that (\ref{equ:wealth_int}) leads to
\begin{equation}\label{equ:wealth_var}
  X_{t+\tau}^f=X_t+\int_t^{t+\tau}(f_s\mu+\alpha)ds+\int_t^{t+\tau}(f_s\sigma)dW_s^{(1)}+\int_t^{t+\tau}\beta dW_s^{(2)},
\end{equation}
for any $\tau>0$.

If we assume that the investment strategy were kept constant during
the time period $(t,t+\tau]$, i.e. $f_s\equiv f$, for any $s\in
(t,t+\tau]$, then it follows immediately from (\ref{equ:wealth_var})
that, given the strategy $f$ and the associated wealth value $X_t=X$
at time $t$, the random variable $\mathcal{X}_{t+\tau}(X,f)$ would
be the future value of the wealth at time $t+\tau$
\begin{equation}\label{equ:benchmark_wealth}
 \mathcal{X}_{t+\tau}(X,f)=X_t+(f\mu+\alpha)\tau+f\sigma(W_{t+\tau}^{(1)}-W_{t}^{(1)})+\beta(W_{t+\tau}^{(2)}-W_{t}^{(2)}).
\end{equation}
Therefore we define the future net worth of the wealth process in
horizon period $(t,t+\tau]$ by $\mathcal{X}_{t+\tau}(X,f)-X_t$.
\newtheorem{def:VaR}{Definition}
\newtheorem{prop:VaR}{Proposition}
\newtheorem{prop:solve_VaR}[prop:VaR]{Proposition}
\newtheorem{Thm:without}{Therom}
\newtheorem{Thm:with}[Thm:without]{Therom}
\begin{def:VaR}[Value-at-risk]
Given a probability level $p\in(0,1)$ and a horizon $\tau>0$, the
value-at-risk of the future net worth of the wealth process with
investment strategy $f$ at time $t$, denoted by $VaR_t^{p,f}$, is
defined as
\begin{equation}
  VaR_t^{p,f}=(Q_t^{p,f})^{-},
\end{equation}
where
\begin{equation}
  Q_t^{p,f}=sup\{L\in \mathbb{R}:P(\mathcal{X}_{t+\tau}-X_t\leq L\mid \mathcal{F}_t)<p\},
\end{equation}
and $x^{-}=max\{0,-x\}.$
\end{def:VaR}

Consequently, $Q_t^{p,f}$ is the $p-$quantile of the projected
portfolio gain over the time interval $(t,t+\tau]$. In other words,
$VaR_t^{p,f}$ is the greatest loss over the next period of length
$\tau$ which would be exceeded only with a small conditional
probability $p$ if the current portfolio $f_t$
were kept the same.

\begin{prop:VaR}[Computation of value-at-risk]\label{prop:VaR}
We have
\begin{equation}\label{equ:VaR}
  VaR_t^{p,f}=(Q_t^{p,f})^{-}=(\tau(f\mu+\alpha)+\sqrt{\tau}\sqrt{f^2\sigma^2+\beta^2+2\rho\sigma\beta f}\Phi^{-1}(p))^{-},
\end{equation}
where $\Phi(\cdot)$ and $\Phi^{-1}(\cdot)$ denote the standard normal
distribution and inverse distribution functions respectively.
\end{prop:VaR}

\begin{proof}[Proof of Proposition \ref{prop:VaR}]
We have
\begin{eqnarray*}
  &&P(\mathcal{X}_{t+\tau}-X_t\leq L\mid \mathcal{F}_t)\\
  &=&P((f\mu+\alpha)\tau+f\sigma(W_{t+\tau}^{(1)}-W_{t}^{(1)})+\beta(W_{t+\tau}^{(2)}-W_{t}^{(2)})\leq L\mid \mathcal{F}_t)\\
  &=&P(f\sigma(W_{t+\tau}^{(1)}-W_{t}^{(1)})+\beta(W_{t+\tau}^{(2)}-W_{t}^{(2)})\leq L-(f\mu+\alpha)\tau\mid \mathcal{F}_t)\\
  &=&P(Z\leq \frac{L-(f\mu+\alpha)\tau}{\sqrt{\tau}\sqrt{f^2\sigma^2+\beta^2+2\rho\sigma\beta f}}\mid \mathcal{F}_t)\\
  &=&\Phi(\frac{L-(f\mu+\alpha)\tau}{\sqrt{\tau}\sqrt{f^2\sigma^2+\beta^2+2\rho\sigma\beta f}}),
\end{eqnarray*}
where $Z=\frac{f\sigma(W_{t+\tau}^{(1)}-W_{t}^{(1)})+\beta(W_{t+\tau}^{(2)}-W_{t}^{(2)})}{\sqrt{\tau}\sqrt{f^2\sigma^2+\beta^2+2\rho\sigma\beta f}}$
follows standard normal distribution conditionally.
Thus, from
\begin{equation*}
  P(\mathcal{X}_{t+\tau}-X_t\leq L\mid \mathcal{F}_t)<p,
\end{equation*}
we know that
\begin{equation*}
  \Phi(\frac{L-(f\mu+\alpha)\tau}{\sqrt{\tau}\sqrt{f^2\sigma^2+\beta^2+2\rho\sigma\beta f}})<p,
\end{equation*}
which gives rise to
\begin{equation*}
  L<\tau(f\mu+\alpha)+\sqrt{\tau}\sqrt{f^2\sigma^2+\beta^2+2\rho\sigma\beta f}\Phi^{-1}(p).
\end{equation*}
And therefore
\begin{eqnarray*}
  Q_t^{p,f}&=&sup\{L\in\mathbb{R}:L<\tau(f\mu+\alpha)+\sqrt{\tau}\sqrt{f^2\sigma^2+\beta^2+2\rho\sigma\beta f}\Phi^{-1}(p)\}\\
  &=&\tau(f\mu+\alpha)+\sqrt{\tau}\sqrt{f^2\sigma^2+\beta^2+2\rho\sigma\beta f}\Phi^{-1}(p),
\end{eqnarray*}
which immediately gives (\ref{equ:VaR}).
\end{proof}

Note that even when $f\equiv0$, which means there is no investment
in risky stock, $VaR_t^{p,0}=\tau(\alpha+\beta\Phi^{-1}(p)/\sqrt{\tau})$ is also
positive if $\alpha/\beta<\Phi^{-1}(1-p)/\sqrt{\tau}$. This results from the
incompleteness of the model, in which the cash flow $Y$ cannot be
traded and therefore the risk cannot be eliminated as long as
$\rho^{2}<1$.\\

\section{Statement of the Problem}

We consider the optimal control problem of the investor who starts
with an initial wealth $X_0=x$ and must select a strategy $f$ so as
to maximize the weighted average of the expectation and the variance
of his terminal wealth, subjected to the constraint that the VaR
with the chosen portfolio is no larger than a given level
$\overline{VaR}$ at any time $t\in[0,T]$. In mathematical terms,
this problem can be described as
\begin{equation}\label{equ:problem}
\begin{array}{c}
\sup\limits_f E[X_T^f-\gamma(X_T^f)^2]\\
\textrm{s.t.}\qquad dX_t=(f_t\mu+\alpha)dt+f_t\sigma dW_t^{(1)}+\beta dW_t^{(2)}, X_0=x,\\
VaR_t^{p,f}\leq\overline{VaR}, \forall t\in[0,T].
\end{array}
\end{equation}
where the parameter (representing the weight) $\gamma$ is positive.
We denote the optimal solution of problem (\ref{equ:problem}) by
$f_{VaR}^{*}$ if it exists. Note that the upper bound
$\overline{VaR}$ can be dependent on $X_t^f$ and $t$, however in
this paper, we set $\overline{VaR}$ to be  a constant in order to
obtain the explicit solution.

\begin{prop:solve_VaR}[Computation of the value-at-risk constraint]
\label{prop:solve_VaR}
The explicit form of the VaR constraint in problem(\ref{equ:problem})
is
\begin{equation}\label{equ:interval}
  \left \{ \begin{array}{ll}

  [\frac{M^2-N^2\beta^2}{2\mu(\rho\beta N-M)},+\infty), & \textrm{if $N\sigma=\mu, \rho\beta N<M$,}\\

  [f_1, +\infty), & \textrm{if $N\sigma<\mu$,}\\

  [f_2, f_1], &  \textrm{if }
  0<N^2\sigma^2-\mu^2\leq\frac{(\sigma M-\rho\beta\mu)^2}{(1-\rho^2)\beta^2},
  \rho\beta\mu<\sigma M,\\

  \emptyset, & \textrm{otherwise},
  \end{array}\right.
\end{equation}
where $N=\Phi^{-1}(1-p)/\sqrt{\tau}>0$, $M=\alpha+\overline{VaR}/\tau>0$,
and
\begin{equation}\label{equ:f_12}
  f_{1,2}=\frac{2(\mu M-\rho\sigma\beta N^2)\pm\sqrt{\Delta}}{2(N^2\sigma^2-\mu^2)},
\end{equation}
if $\Delta=4N^2((1-\rho^2)\beta^2(\mu^2-N^2\sigma^2)+(\sigma M-\rho\beta\mu)^2)\geq0$
and $N^2\sigma^2\neq\mu^2$. We also assume that $f_1$ is
always larger than $f_2$.
\end{prop:solve_VaR}

\begin{proof}[Proof of Proposition \ref{prop:solve_VaR}]
The VaR constraint can be written as
\begin{equation*}
  \tau(f\mu+\alpha+\frac{\Phi^{-1}(p)}{\sqrt{\tau}}\sqrt{f^2\sigma^2+\beta^2+2\rho\sigma\beta f})^{-}\leq\overline{VaR},
\end{equation*}
which is equivalent to
\begin{equation*}
  \left \{\begin{array}{l}
  (\frac{\Phi^{-1}(p)}{\sqrt{\tau}})^2(f^2\sigma^2+\beta^2+2\rho\sigma\beta f)\leq(f\mu+\alpha+\frac{\overline{VaR}}{\tau})^2,\\
  f\mu+\alpha+\frac{\overline{VaR}}{\tau}\geq0,
  \end{array}\right.
\end{equation*}
that is, after some simplifications,
\begin{equation*}
  \left \{\begin{array}{l}
  N^2(f^2\sigma^2+\beta^2+2\rho\sigma\beta f)\leq(f\mu+M)^2,\\
  f\mu+M\geq0,
  \end{array}\right.
\end{equation*}
where $N=\Phi^{-1}(1-p)/\sqrt{\tau}>0$, $M=\alpha+\overline{VaR}/\tau>0$.
Note that when $f=-M/\mu$,
$ N^2(f^2\sigma^2+\beta^2+2\rho\sigma\beta f)>0=(f\mu+M)^2$ always holds.
This observation will help us in the second case when $N^2\sigma^2-\mu^2<0$
and the third case when $N^2\sigma^2-\mu^2>0$.

Therefore we have a group of inequalities
\begin{equation}\label{equ:constraint}
\left \{\begin{array}{l}
  (N^2\sigma^2-\mu^2)\cdot f^2+2(\rho\sigma\beta N^2-\mu M)\cdot f+(N^2\beta^2-M^2)\leq0,\\
  f\geq-M/\mu.
  \end{array}\right.
\end{equation}

First we study the degeneration case: $N^2\sigma^2-\mu^2=0$,i.e.
$N\sigma=\mu$.

In this case, if $2(\rho\sigma\beta N^2-\mu M)>0$, i.e.
$\rho\beta N>M$, then the first inequality of
(\ref{equ:constraint}) would imply
\begin{displaymath}
f\leq \frac{M^2-N^2\beta^2}{2(\rho\sigma\beta N^2-\mu M)}
=\frac{M^2-N^2\beta^2}{2\mu(\rho\beta N-M)},
\end{displaymath}
however
\begin{displaymath}
  -\frac{M}{\mu}>\frac{M^2-N^2\beta^2}{2\mu(\rho\beta N-M)},
\end{displaymath}
which, together with the second inequality of (\ref{equ:constraint}), leads to $f\in \emptyset$.

If $2(\rho\sigma\beta N^2-\mu M)=0$, i.e.
$\rho\beta N=M$, then the first inequality of
(\ref{equ:constraint}) would imply
\begin{displaymath}
N^2\beta^2-M^2=N^2\beta^2(1-\rho^2)\leq0,
\end{displaymath}
which leads to a contradiction. Thus there is no
solution.

If $2(\rho\sigma\beta N^2-\mu M)<0$, i.e.
$\rho\beta N<M$, then the first inequality of
(\ref{equ:constraint}) would imply
\begin{displaymath}
f\geq\frac{M^2-N^2\beta^2}{2\mu(\rho\beta N-M)},
\end{displaymath}
so the constraint becomes
\begin{displaymath}
f\in[max\{-\frac{M}{\mu}, \frac{M^2-N^2\beta^2}{2\mu(\rho\beta N-M)}\},+\infty)
=[\frac{M^2-N^2\beta^2}{2\mu(\rho\beta N-M)},+\infty).
\end{displaymath}

Secondly, we study the case: $N^2\sigma^2-\mu^2<0$.

In this case,
$\Delta=4N^2((1-\rho^2)\beta^2(\mu^2-N^2\sigma^2)+(\sigma M-\rho\beta\mu)^2)>0$.
$f_2<f_1$ are two points of intersection of the parabola and
the lateral axis. Using the observation we have
mentioned just now, we know that the first inequality
of (\ref{equ:constraint}) will not hold if
$f=-M/\mu$, which leads to $f_2<-M/\mu<f_1$.
Therefore in this case, the constraint is
$f\in[f_1, +\infty)$.

Thirdly, we study the case: $N^2\sigma^2-\mu^2>0$.

In this case, if $\Delta\geq0$, i.e.
\begin{displaymath}
N^2\sigma^2-\mu^2\leq\frac{(\sigma M-\rho\beta\mu)^2}{(1-\rho^2)\beta^2}.
\end{displaymath}
The constraint becomes $f\in[f_2, f_1]\cap[-M/\mu, +\infty)$.
Also by using the observation, we have $-M/\mu<f_2\leq f_1$
or $f_2\leq f_1<-M/\mu$. And therefore we have
\begin{displaymath}
f\in[f_2, f_1], \textrm{if $-\frac{M}{\mu}<\frac{\mu M-\rho\sigma\beta N^2}{N^2\sigma^2-\mu^2}$},
\end{displaymath}
or
\begin{displaymath}
f\in\emptyset, \textrm{if $-\frac{M}{\mu}>\frac{\mu M-\rho\sigma\beta N^2}{N^2\sigma^2-\mu^2}$},
\end{displaymath}
where $(\mu M-\rho\sigma\beta N^2)/(N^2\sigma^2-\mu^2)=\frac{1}{2}(f_1+f_2)$
is the symmetry axis of the parabola.

If $\Delta<0$, then the parabola is strictly above
the lateral axis, and thus no solution can satisfy
even the first inequality of (\ref{equ:constraint}).
\end{proof}

\section{Optimal Mean-variance Strategy without VaR Constraints}

Before we finally solve the problem (\ref{equ:problem}),  we
consider the corresponding optimal control problem without the VaR
constraint
\begin{equation}\label{equ:problem_without}
\begin{array}{c}
\sup\limits_f E[X_T^f-\gamma(X_T^f)^2]\\
\textrm{s.t.}\qquad dX_t=(f_t\mu+\alpha)dt+f_t\sigma
dW_t^{(1)}+\beta dW_t^{(2)}, X_0=x.
\end{array}
\end{equation}
Here we denote the optimal solution of problem
(\ref{equ:problem_without}) by $f^*$.

To proceed, let $V(t,x)=\sup\limits_f E[u(X_T^f)|X_t^f=x]$ be the
optimal value function attainable by the investor starting from the
state $x$ at time $t$. And we will give the explicit form of the
optimal strategy $f^*$ in the following theorem.

\begin{Thm:without}[The optimal mean-variance strategy without VaR constraints]\label{Thm:without}
The optimal strategy to maximize expected utility at
terminal time $T$ is to invest, at each time $t\leq T$,
\begin{equation}\label{equ:f^**}
f^*=-\frac{\mu}{\sigma^2}(x-\frac{1}{2\gamma})
-\frac{\mu(\alpha-\rho\beta\mu/\sigma)(T-t)}{\mu^2(T-t)+\sigma^2}-\frac{\rho\beta}{\sigma},
\end{equation}
and then the optimal value function is
\begin{equation}\label{equ:Valuee}
V(t,x)=-\gamma(x-\frac{1}{2\gamma})^2e^{k_1(T-t)}+k_2(T-t)(x-\frac{1}{2\gamma})
+k_3(T-t)+\frac{1}{4\gamma},
\end{equation}
where
\begin{equation}\label{equ:k1k2k3}
\left \{\begin{array}{l}
k_1=2B,\\
k_2=\frac{2A\gamma e^{2B(T-t)}}{2B(T-t)-1},\\
k_3=2A^2\gamma e^{2B(T-t)}[\frac{T-t}{2B(T-t)-1}-B(\frac{T-t}{2B(T-t)-1})^2-\frac{C}{A^2}],
\end{array}\right.
\end{equation}
and
$A=\alpha-\rho\beta\mu/\sigma$,
$B=-\frac{1}{2}(\mu/\sigma)^2$,
$C=\frac{1}{2}\beta^2(1-\rho^2)$.
\end{Thm:without}

\begin{proof}
From Fleming and Rishel (1975), the corresponding HJB equation is
given by
\begin{equation}\label{equ:HJB}
\left \{ \begin{array}{l}
\sup\limits_f\{V_t +[f\mu+\alpha]V_x+\frac{1}{2}[f^2\sigma^2+\beta^2+2\rho\sigma\beta f]V_{xx}\}=0,\\
V(T,x)=x-\gamma x^2.
\end{array}\right.
\end{equation}
Assume that the HJB equation (\ref{equ:HJB}) has a classic solution
$V$, which satisfies $V_{xx}<0$. Then differentiating with respect
to $f$ gives the optimizer
\begin{equation}\label{equ:f^*}
  f^*=-\frac{\mu}{\sigma^2}\frac{V_x}{V_{xx}}-\frac{\rho\beta}{\sigma}.
\end{equation}
Substituting (\ref{equ:f^*}) back into (\ref{equ:HJB}), the HJB
equation becomes, after some simplification, equivalent to the
following nonlinear Cauchy problem for the value function $V$
\begin{equation}\label{equ:cauchy}
  \left \{ \begin{array}{l}
  V_t+AV_x+B\frac{V_x^2}{V_{xx}}+CV_{xx}=0,\\
  V(T,x)=x-\gamma x^2=-\gamma(x-\frac{1}{2\gamma})^2+\frac{1}{4\gamma},
  \end{array}\right.
\end{equation}
where the constants $A=\alpha-\rho\beta\mu/\sigma$,
$B=-\frac{1}{2}(\mu/\sigma)^2$,
$C=\frac{1}{2}\beta^2(1-\rho^2)$.

In order to simplify the boundary condition, let
\begin{equation}
  y=x-\frac{1}{2\gamma},
\end{equation}
and rewrite the HJB equation (\ref{equ:cauchy}) into the form
\begin{equation}\label{equ:cauchyy}
  \left \{ \begin{array}{l}
  V_t+AV_y+B\frac{V_y^2}{V_{yy}}+CV_{yy}=0,\\
  V(T,y)=-\gamma y^2+\frac{1}{4\gamma}.
  \end{array}\right.
\end{equation}

To solve this partial differential equation (\ref{equ:cauchyy}), we
try to fit a solution of the form
\begin{equation}\label{equ:solution}
  V(t,y)=-\gamma y^2e^{k_1(T-t)}+k_2(T-t)y+k_3(T-t)+\frac{1}{4\gamma},
\end{equation}
where $k_1$, $k_1$ and $k_1$ are suitable coefficient,
and note that by the form of (\ref{equ:solution}) we
have
  \begin{eqnarray}
  V_t&=&k_1\gamma y^2e^{k_1(T-t)-k_2y-k_3},\nonumber\\
  V_y&=&-2\gamma e^{k_1(T-t)}y+k_2(T-t),\label{equ:Value}\\
  V_{yy}&=&-2\gamma e^{k_1(T-t)}\nonumber.
  \end{eqnarray}
The boundary condition is naturally satisfied by the solution form
of (\ref{equ:solution}). By substituting (\ref{equ:Value}) into
(\ref{equ:cauchyy}), we have
\begin{equation}
\begin{split}
  (k_1\gamma y^2e^{k_1(T-t)}-k_2y-k_3)
  +A(-2\gamma e^{k_1(T-t)}y+k_2(T-t))\\
  +B\frac{[-2\gamma e^{k_1(T-t)}y+k_2(T-t)]^2}{-2\gamma e^{k_1(T-t)}}
  +C(-2\gamma e^{k_1(T-t)})=0,
  \end{split}
\end{equation}
which requires
$k_1$, $k_2$ and $k_3$ to satisfy
\begin{equation}\label{equ:kkk}
  \left \{ \begin{array}{l}
  k_1\gamma e^{k_1(T-t)}+B(-2\gamma e^{k_1(T-t)})=0,\\
  -k_2-2A\gamma e^{k_1(T-t)}+2Bk_2(T-t)=0,\\
  -k_3+Ak_2(T-t)+B\frac{k_2^2(T-t)^2}{-2\gamma e^{k_1(T-t)}}+C(-2\gamma e^{k_1(T-t)})=0.
  \end{array}\right.
\end{equation}
Solving equations (\ref{equ:kkk}), we derive (\ref{equ:k1k2k3}), and
hence (\ref{equ:Valuee}) after replacing $y$ by
$(x-\frac{1}{2\gamma})$. Since we have the value function in
explicit form, it becomes easy for us to obtain the optimal control
of (\ref{equ:f^**}) by substituting the value for $V_x$ and $V_{xx}$
from (\ref{equ:Value}) into (\ref{equ:f^*}).

\end{proof}

\section{Optimal Mean-variance Strategy under VaR Constraints}\label{sec:with_VaR}

In this section, we come back to the problem (\ref{equ:problem}).
Our objective is to give the optimal control $f_{VaR}^*$ as well as
the corresponding value function $V$ in explicit form. Since in
Theorem~\ref{Thm:without} we have already found the optimal control
$f^*$ which optimize the problem of (\ref{equ:problem_without})
without the VaR constraint Therefore, we obtain $f_{VaR}^*=f^*$ as
long as $f^*$ satisfies the VaR constraint. However, the global
optimizer can not be the local optimizer when $f^*$ fails the VaR
constraint.

Remind ourselves of the proof in Theorem \ref{Thm:without}, by
applying the dynamic programming approach we are tackling with a
static optimization problem (\ref{equ:HJB}). Rewrite the problem and
add the VaR constraint to it and we have
\begin{equation}
\begin{array}{c}
\sup\limits_f\{\frac{1}{2}V_{xx}\sigma^2\cdot f^2
+(\mu V_x+\rho\beta\sigma V_{xx})\cdot f+(V_t+\alpha V_x
+\frac{1}{2}\beta^2V_{xx})\},\\
\textrm{s.t. $f\in I$}
\end{array}
\end{equation}
where the intervals $I$ have the same forms of (\ref{equ:interval})
in Proposition~\ref{prop:solve_VaR}, therefore, the constraint $f\in
I$ is equivalent to the VaR constraint in problem
(\ref{equ:problem}). Since we know that
$\frac{1}{2}V_{xx}\sigma^2<0$, it becomes a simple problem to find
the peak of the parabola, which opens down, on each interval $I$.
Specifically, \begin{description}

\item[Case~1]
when $N\sigma=\mu, \rho\beta N<M$,
\begin{equation}
  f_{VaR}^*=\left \{ \begin{array}{ll}
  f^*, & f^*\in[\frac{M^2-N^2\beta^2}{2\mu(\rho\beta N-M)},+\infty),\\
  \frac{M^2-N^2\beta^2}{2\mu(\rho\beta N-M)}, & \textrm{otherwise}.
  \end{array}\right.
\end{equation}

\item[Case~2]
when $N\sigma<\mu$,
\begin{equation}\label{equ:f_VaR_30}
  f_{VaR}^*=\left \{ \begin{array}{ll}
  f^*, & f^*\in[f_1,+\infty),\\
  f_1, & \textrm{otherwise}.
  \end{array}\right.
\end{equation}

\item[Case~3]
when $0<N^2\sigma^2-\mu^2\leq\frac{(\sigma M-\rho\beta\mu)^2}{(1-\rho^2)\beta^2},
\rho\beta\mu<\sigma M$,
\begin{equation}\label{equ:f_VaR_40}
  f_{VaR}^*=\left \{ \begin{array}{ll}
  f^*, & f^*\in[f_2, f_1],\\
  f_2, & f^*\in(-\infty, f_2),\\
  f_1, & f^*\in(f_1, +\infty).
  \end{array}\right.
\end{equation}
\end{description}

Except these three cases listed above, there is
no solution which can satisfies the constraint.
Here $f_1$ and $f_2$ have the same expressions
as shown in Proposition~\ref{prop:solve_VaR},
and $f^*$ has the expression of (\ref{equ:f^**})
in Theorem~\ref{Thm:without}.

Before giving the optimal control as well as the corresponding value
function in explicit form, we have to admit that we are going to
omit the first case, i.e. when $N\sigma=\mu$ happens. On the one
hand, this case could hardly happen so that we can benefit little
from studying them in practice,
on the other hand, the procedures of studying the first case is
quite similar with the other two, it will be therefore a mere
repetition.

\begin{Thm:with}[The optimal mean-variance strategy under VaR constraints]
\label{Thm:with}
The optimal strategy to maximize expected utility
at terminal time $T$ subjected to the VaR constraint
is to invest $f_{VaR}^*$.

When $N\sigma<\mu$, \begin{equation}\label{equ:f_VaR_3}
f_{VaR}^*=\max\{f_1, f^*\},
\end{equation}
and the optimal value function is
\begin{equation}\label{equ:Value_3}
  V(t,x)=\left \{ \begin{array}{ll}
  \begin{split}-\gamma(x-\frac{1}{2\gamma})^2e^{k_1(T-t)}+k_2(T-t)(x-\frac{1}{2\gamma})\\
  +k_3(T-t)+\frac{1}{4\gamma}, \end{split}& f^* \geq f_1,\\
  \begin{split}\frac{1}{4\gamma}-\gamma[(x-\frac{1}{2\gamma})^2
  +2D_1(x-\frac{1}{2\gamma})(T-t)\\
  +D_1^2(T-t)^2+2E_1(T-t)], \end{split}& f^* < f_1;
  \end{array}\right.
\end{equation}

When $0<N^2\sigma^2-\mu^2\leq\frac{(\sigma M-\rho\beta\mu)^2}{(1-\rho^2)\beta^2},
\rho\beta\mu<\sigma M$,
\begin{equation}\label{equ:f_VaR_4}
  f_{VaR}^*=\max\{f_2, \min\{f_1, f^*\}\},
\end{equation}
and the optimal value function is
\begin{equation}\label{equ:Value_4}
  V(t,x)=\left \{ \begin{array}{ll}
  \begin{split}-\gamma(x-\frac{1}{2\gamma})^2e^{k_1(T-t)}+k_2(T-t)(x-\frac{1}{2\gamma})\\
  +k_3(T-t)+\frac{1}{4\gamma}, \end{split}& f^* \in [f_2, f_1],\\

  \begin{split}\frac{1}{4\gamma}-\gamma[(x-\frac{1}{2\gamma})^2
  +2D_2(x-\frac{1}{2\gamma})(T-t)\\
  +D_2^2(T-t)^2+2E_2(T-t)], \end{split}& f^* \in (-\infty, f_2),\\

  \begin{split}\frac{1}{4\gamma}-\gamma[(x-\frac{1}{2\gamma})^2
  +2D_1(x-\frac{1}{2\gamma})(T-t)\\
  +D_1^2(T-t)^2+2E_1(T-t)], \end{split}& f^* \in (f_1, +\infty)
  \end{array}\right.
\end{equation}
where $k_1$, $k_2$ and $k_3$ have the same expressions
of (\ref{equ:k1k2k3}) in Theorem~\ref{Thm:without}, $f_{1,2}$
have the expressions of (\ref{equ:f_12}), and $D_i=f_i\mu+\alpha$,
$E_i=\frac{1}{2}(f_i^2\sigma^2+\beta^2+2\rho\sigma\beta f_i)$,
for $i=1,2$.

\end{Thm:with}

\begin{proof}
Since the explicit form (\ref{equ:f_VaR_3}) and (\ref{equ:f_VaR_4})
of the optimal control $f_{VaR}^*$ in this theorem are only the
rescript of (\ref{equ:f_VaR_30}) and (\ref{equ:f_VaR_40}), we only
have to work out with the corresponding value function $V$.

When $N\sigma<\mu$ happens, substituting (\ref{equ:f_VaR_30}) in
(\ref{equ:HJB}) and we have
\begin{equation}\label{equ:cauchy_3}
  0=\left \{ \begin{array}{ll}
  V_t+(\alpha-\rho\beta\mu/\sigma)V_x-\frac{1}{2}(\mu/\sigma)^2\frac{V_x^2}{V_{xx}}
  +\frac{1}{2}\beta^2(1-\rho^2)V_{xx},  & f^*\geq f_1,\\
  V_t+(f_1\mu+\alpha)V_x+\frac{1}{2}(f_1^2\sigma^2+\beta^2+2\rho\sigma\beta f_1)V_{xx}, & f^*<f_1,
  \end{array}\right.
\end{equation}
with the same terminal condition $V(T,x)=x-\gamma x^2$.

Since we have already solved the first equation of
(\ref{equ:cauchy_3}) in Theorem~\ref{Thm:without},
we will focus on the second equation which appears
to be much more easier to handle for its linearity.

Using the notation of $D_1$ and $E_1$, the second
equation of (\ref{equ:cauchy_3}) with the terminal
condition can be written as another Cauchy problem
\begin{equation}
  \left \{ \begin{array}{l}
  V_t+D_1V_x+E_1V_{xx}=0, \textrm{for $t<T$},\\
  V(T,x)=x-\gamma x^2.
  \end{array} \right.
\end{equation}
Applying the same trick by letting $y=x-\frac{1}{2\gamma}$,
then
\begin{equation}
  \left \{ \begin{array}{l}
  V_t+D_1V_y+E_1V_{yy}=0, \textrm{for $t<T$},\\
  V(T,y)=-\gamma y^2+\frac{1}{4\gamma}.
  \end{array}. \right.
\end{equation}
In order to apply the Fourier transform, change
the terminal condition into initial condition
by letting $U(t, y)=V(T-t, y)$
\begin{equation}
  \left \{ \begin{array}{l}
  U_t-D_1U_y-E_1U_{yy}=0, \textrm{for $t>0$},\\
  U(0,y)=-\gamma y^2+\frac{1}{4\gamma}.
  \end{array} \right.
\end{equation}
Then we have
\begin{equation}
  U(t,y)=\gamma(\frac{1}{4\gamma^2}-y^2-2D_1yt-D_1^2t^2-2E_1t),
\end{equation}
which immediately gives the second part of (\ref{equ:Value_3}).

Similarly when $0<N^2\sigma^2-\mu^2\leq\frac{(\sigma M-\rho\beta\mu)^2}{(1-\rho^2)\beta^2},
\rho\beta\mu<\sigma M$, the corresponding
Cauchy problem is described as
\begin{equation}\label{equ:cauchy_4}
  0=\left \{ \begin{array}{ll}
  V_t+(\alpha-\rho\beta\mu/\sigma)V_x-\frac{1}{2}(\mu/\sigma)^2\frac{V_x^2}{V_{xx}}
  +\frac{1}{2}\beta^2(1-\rho^2)V_{xx},  & f^*\in[f_2, f_1],\\
  V_t+(f_2\mu+\alpha)V_x+\frac{1}{2}(f_2^2\sigma^2+\beta^2+2\rho\sigma\beta f_2)V_{xx}, & f^*\in(-\infty, f_2),\\
  V_t+(f_1\mu+\alpha)V_x+\frac{1}{2}(f_1^2\sigma^2+\beta^2+2\rho\sigma\beta f_1)V_{xx}, & f^*\in(f_1, +\infty),
  \end{array}\right.
\end{equation}
with terminal condition $V(T,x)=x-\gamma x^2$.

Almost the same steps can be taken if we replace
the notation $f_1$, $D_1$ and $E_1$ with $f_2$, $D_2$ and $E_2$
respectively, and (\ref{equ:Value_4}) will be
obtained.

\end{proof}

\section{Illustration of the Solutions}

In this section we will illustrate the result of
Theorem~\ref{Thm:without} and Theorem~\ref{Thm:with}.
Without loss of generality, we set the initial wealth
level between $[0,1]$. The VaR horizon period is chosen to be
1 trading day, nearly 1/260 calendar year, while the
terminal year is set to be 10 calendar year. Confidence
level is $1-p=99\%$, and the upper VaR limit is
$\overline{VaR}=0.02$, which is $2\%$ of the initial
wealth. For the stochastic cash flow, we use $\alpha=0.01$,
$\beta=0.14$. In the market, the risk-free interest
$r=0$, as we always assumed, and for the stock,
$\mu=0.05$, $\sigma=0.3$. the correlation coefficient
of $W_t^{(1)}$ and $W_t^{(2)}$ is set to be $\rho=0.2$,
and the parameter in the quadratic utility function is
$\gamma=1$. We summarize these parameters below in
the Table~\ref{tab:para_4}.
\begin{table}[htb]
\caption{Parameters} \label{tab:para_4}

\newcommand{\m}{\hphantom{$-$}}
\newcommand{\cc}[1]{\multicolumn{1}{c}{#1}}
\renewcommand{\tabcolsep}{0.8pc} 
\renewcommand{\arraystretch}{1.2} 

\begin{tabular}{@{}lllllllllll}
\hline
$x$ & $\mu$ & $\sigma$ & $\alpha$ & $\beta$ & $\rho$ & $\gamma$ & $T$ & $\tau$ & $p$ & $\overline{VaR}$\\
\hline
1   & 0.05  &   0.3    &   0.01   &  0.14   &  0.2   &  1       &  10 & 1/260  & 0.01&   0.02    \\
\hline
\end{tabular}
\end{table}

The setting of the parameters in Table~\ref{tab:para_4}
satisfies the three conditions of case~3:
\begin{displaymath}
0<N^2\sigma^2-\mu^2\leq\frac{(\sigma M-\rho\beta\mu)^2}{(1-\rho^2)\beta^2},
\rho\beta\mu<\sigma M,
\end{displaymath}
where $N=N^{-1}(1-p)/\sqrt{\tau}=37.74$,
$M=\alpha+\overline{VaR}/\tau=5.273$. Therefore the
optimal control has the expression
$f_{VaR}^*=\max\{f_2, \min\{f_1, f^*\}\}$. We show
the optimal strategy without the VaR limit as well
as the constrained one on $(x,t)\in[0,1]\times[0,10]$
in the Figure~\ref{fig:f_4} below.
\begin{figure}[H]
\centering
  \includegraphics[width=9cm]{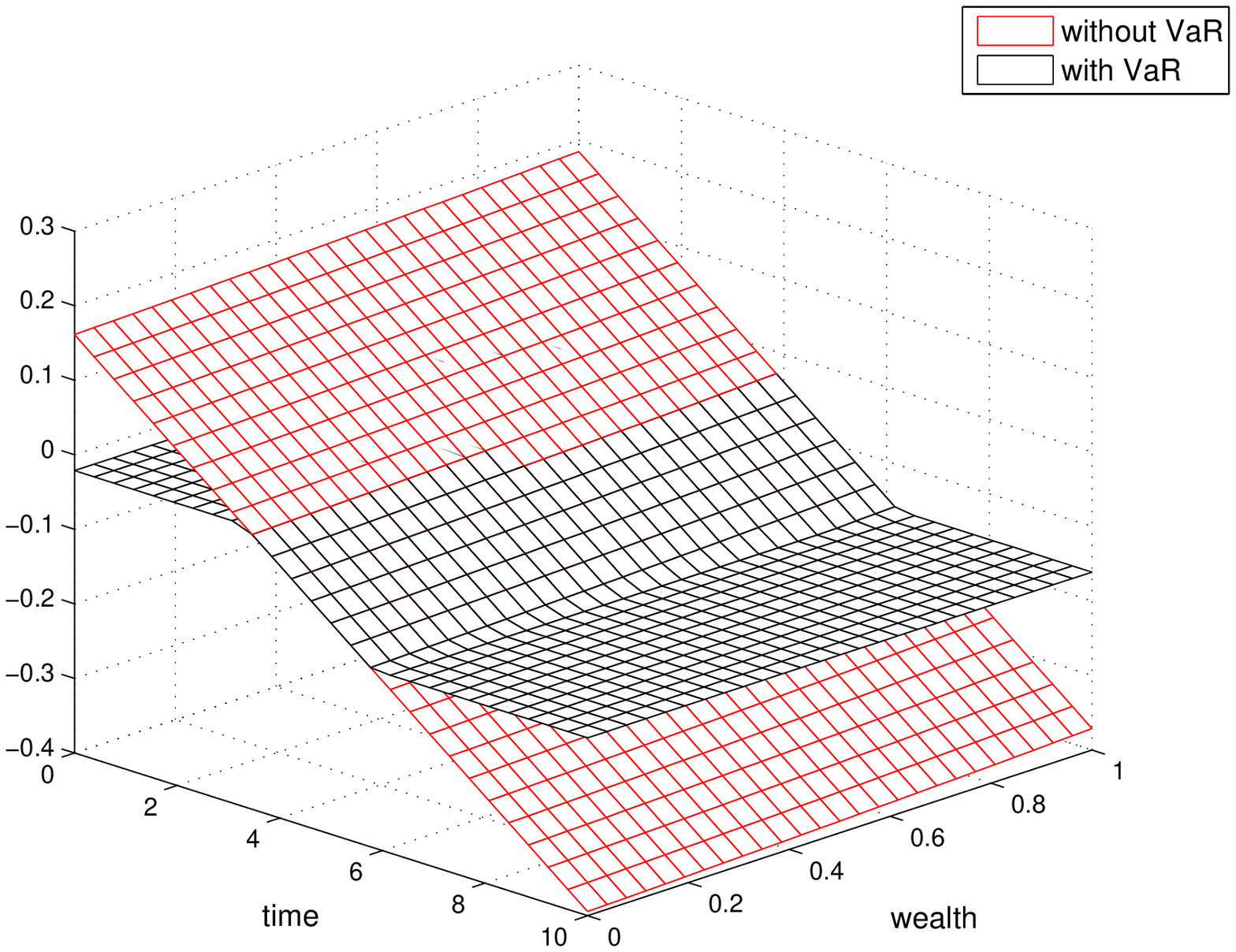}
  \caption{Optimal strategies in case~3}
  \label{fig:f_4}
\end{figure}

As shown in this figure, the VaR constraint
actually gives an upper and a lower bound surface to
the strategy surface. From the expression of (\ref{equ:f_12}),
we know that $f_{1,2}$ are independent of $x$ and $t$,
and strictly negative in this case. Therefore the bound
surfaces are horizontal and below level zero, which
actually constrain the behavior of short-selling.

\begin{figure}[H]
\centering
  \includegraphics[height=8cm]{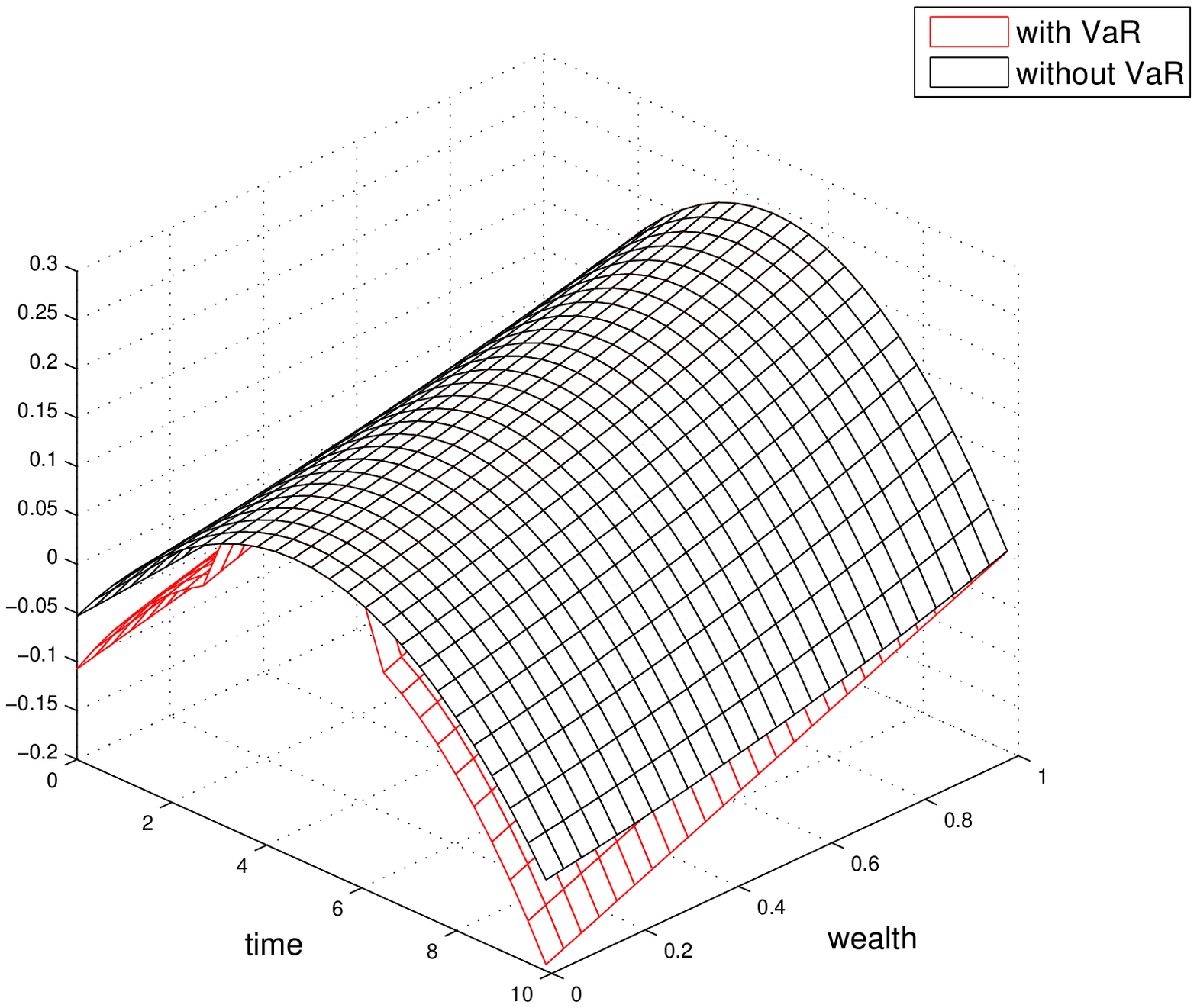}
  \caption{Value functions in case~3}
  \label{fig:value_4}
\end{figure}

We could also compare the optimal value function with and without
the VaR constraint, which are illustrated in
Figure~\ref{fig:value_4} for case 3. In both figures we could
observe that the VaR constraint is active during the time period of
approximately $t\in[0,3]\cup[6,10]$. The optimal function of
constrained problem is identical to that of the unconstrained one
during $t\in[3,6]$, and it becomes inferior when the constrain is
active.

By changing the value of drift and diffusion
parameter of the risky stock, investment in
them becomes much more promising than in
case~3. See the table below, note that we
do not change the values of other parameters.
\begin{table}[htb]
\caption{Parameters} \label{tab:para_3}

\newcommand{\m}{\hphantom{$-$}}
\newcommand{\cc}[1]{\multicolumn{1}{c}{#1}}
\renewcommand{\tabcolsep}{0.8pc} 
\renewcommand{\arraystretch}{1.2} 

\begin{tabular}{@{}lllllllllll}
\hline
$x$ & $\mu$ & $\sigma$ & $\alpha$ & $\beta$ & $\rho$ & $\gamma$ & $T$ & $\tau$ & $p$ & $\overline{VaR}$\\
\hline
1   & 0.8   &   0.02   &   0.01   &  0.14   &  0.2   &  1       &  10 & 1/260  & 0.01&   0.02    \\
\hline
\end{tabular}
\end{table}

\begin{figure}[H]
\centering
  \includegraphics[height=8cm]{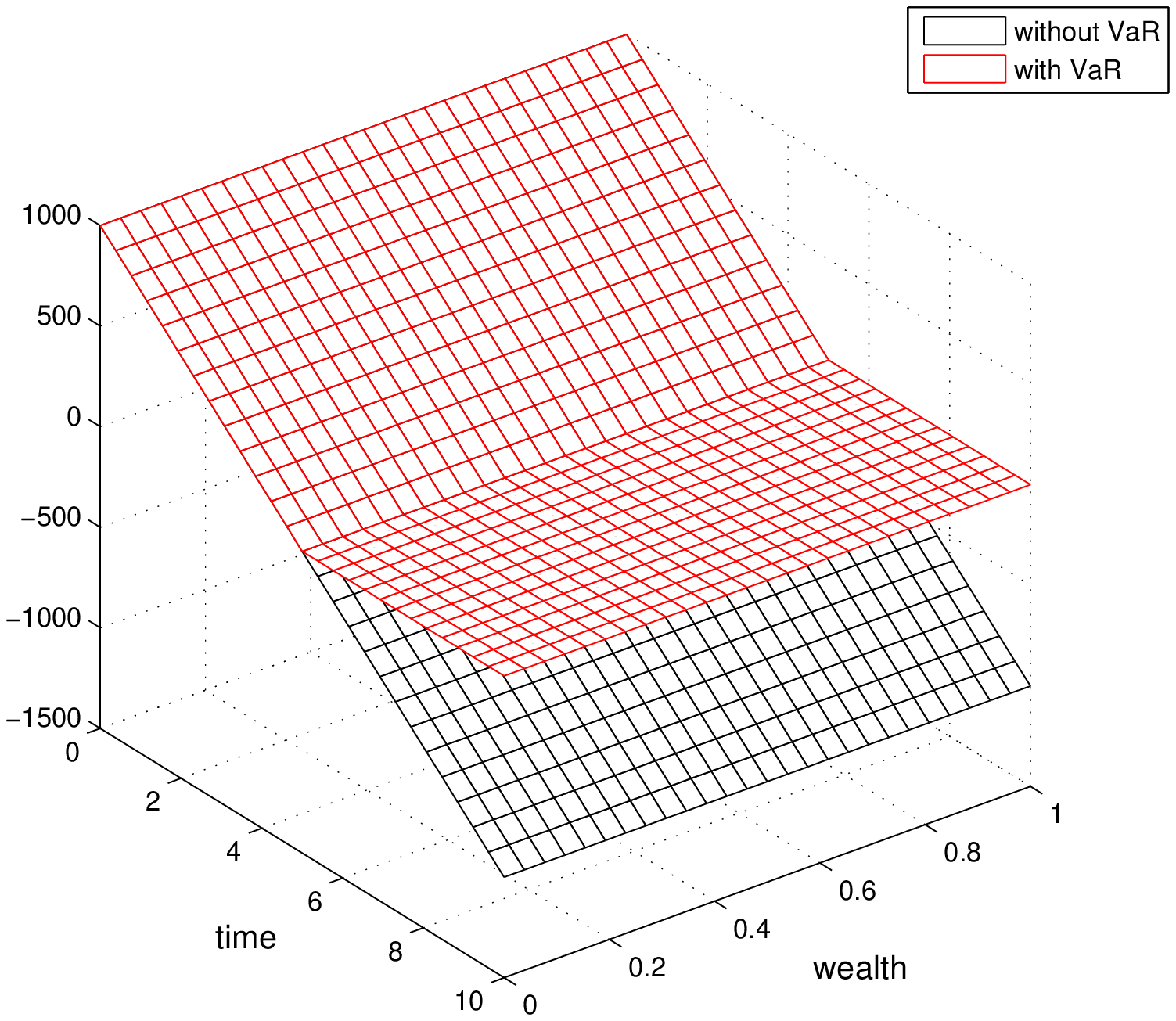}
  \caption{Optimal strategies in case~2}
  \label{fig:f_3}
\end{figure}

\begin{figure}[H]
\centering
  \includegraphics[height=8cm]{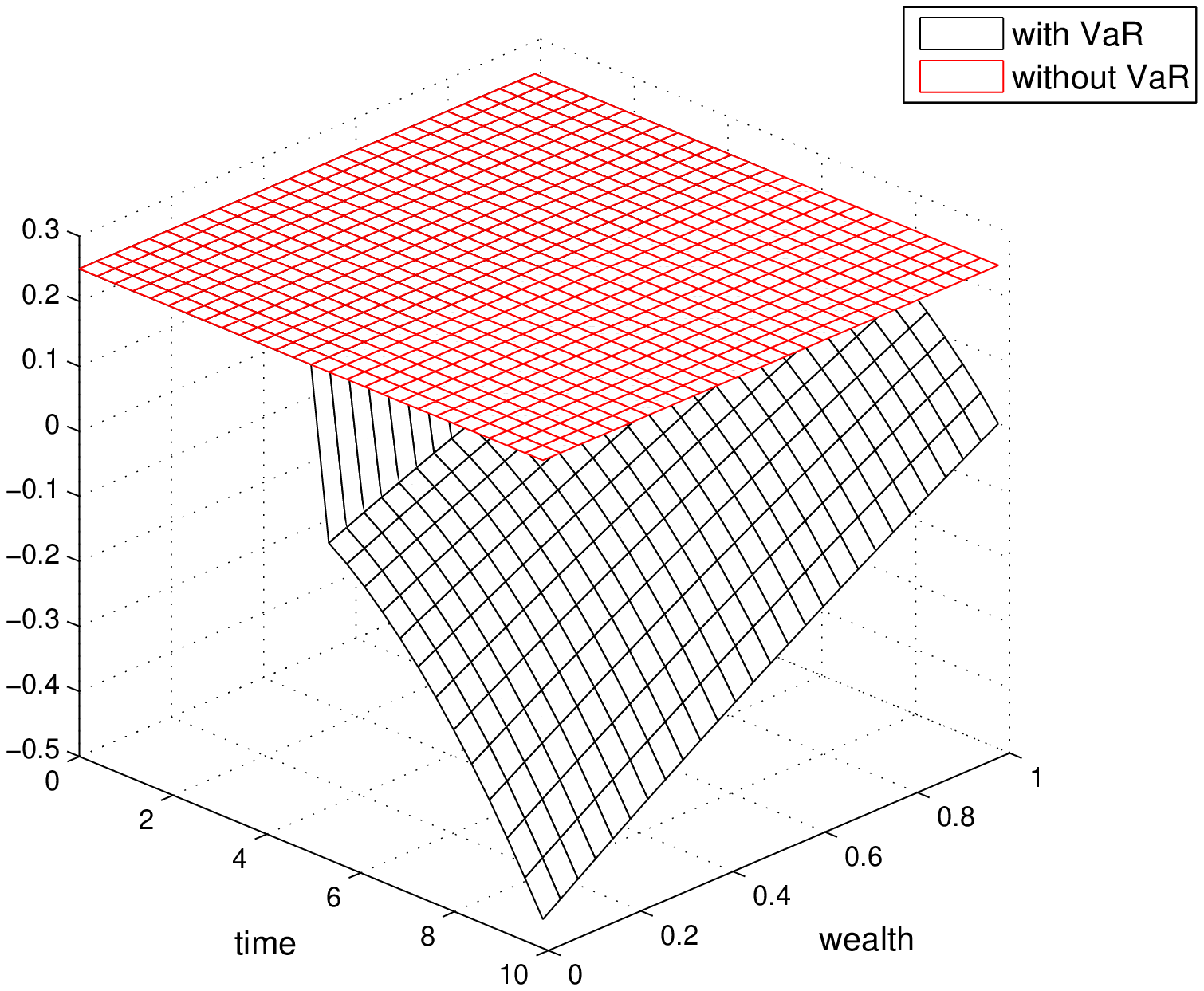}
  \caption{Value functions in case~2}
  \label{fig:value_3}
\end{figure}

The setting of the parameters in Table~\ref{tab:para_3}
satisfies the condition of case~2, i.e. $N\sigma<\mu$,
where $N=37.74$. In Figure~\ref{fig:f_3}, we observe
that when time $t$ is near 0, the investment in
risky asset appears to be rather radical. This results
from the superiority of the stock, and the investor
can hardly get any loss under such circumstance, and
therefore do not activate the VaR constraint. When
time goes to approximately $t=5$, the constraint
becomes active and gives the optimal strategy a
lower bound near level zero, which is the only bound
surface brought by the VaR constraint in case~2.
The optimal value function surfaces are shown in
Figure~\ref{fig:value_3}, where the surface of the
constrained problem remains identical to that of
the unconstrained problem until $t=5$. All as we
expected, the constrained surface becomes inferior
since then. Note that in this case, the optimal
value without VaR constraint appears almost
horizontal, which seems unreasonable. This results
from the uncommon condition that $N<\mu/\sigma$,
which directly leads to the huge absolute value of
$B=-0.5(\mu/\sigma)^2$. Look at the expressions of
(\ref{equ:Valuee}) and (\ref{equ:k1k2k3}), the
factor $e^{2B(T-t)}$ almost annihilate the first
three terms in $V(t,x)$, which makes $V(t,x)$ becomes
a constant of $1/(4\gamma)$ approximately.

\bibliographystyle{model2-names}
\bibliography{refs}







\end{document}